\documentclass[aps,pra,reprint,showpacs,superscriptaddress,nofootinbib]{revtex4-1}
\usepackage[utf8]{inputenc}
\usepackage[english]{babel}

\usepackage[
	unicode           = true  ,
	plainpages        = false , 
	pdfpagelabels     = true  , 
	bookmarks         = true  ,
	bookmarksnumbered = true  ,
	bookmarksopen     = true  ,
	breaklinks        = true  ,
	backref           = false ,
	colorlinks        = true  ,
	linkcolor         = blue  ,
	urlcolor          = blue  ,
	citecolor         = red   ,
	anchorcolor       = green ,
	hyperindex        = true  ,
	linktocpage       = true  ,
	hyperfigures      = true
]{hyperref}
\hypersetup{
    pdftitle={Theory of chiral edge state lasing in a two-dimensional topological system},
    pdfauthor={Matteo Seclì <matteo.secli@sissa.it>}
}

\usepackage{amsmath}
\usepackage{amssymb}

\usepackage{graphicx}
\usepackage[export]{adjustbox}
\graphicspath{{figures/PNG/}{figures/PDF/}{figures/}}

\begin{document}

\title{Theory of chiral edge state lasing in a two-dimensional topological system}

\author{Matteo Seclì}
\email[]{matteo.secli@sissa.it}
\affiliation{International School for Advanced Studies (SISSA), Via Bonomea 265, I-34136 Trieste, Italy}

\author{Massimo Capone}
\affiliation{International School for Advanced Studies (SISSA), Via Bonomea 265, I-34136 Trieste, Italy}
\affiliation{CNR-IOM Democritos, Via Bonomea 265, I-34136 Trieste, Italy}

\author{Iacopo Carusotto}
\email[]{iacopo.carusotto@unitn.it}
\affiliation{INO-CNR BEC Center and Dipartimento di Fisica, Università di Trento, I-38123 Povo, Italy}

\date{\today}

\begin{abstract}
We theoretically study topological laser operation in a bosonic Harper-Hofstadter model featuring a saturable optical gain. Crucial consequences of the chirality of the lasing edge modes are highlighted, such as a sharp dependence of the lasing threshold on the geometrical shape of the amplifying region and the possibility of ultraslow relaxation times and of convectively unstable regimes. The different unstable regimes are characterized in terms of spatio-temporal structures sustained by noise and a strong amplification of a propagating probe beam is anticipated to occur in between the convective and the absolute (lasing) thresholds. The robustness of topological laser operation against static disorder is assessed. 
\end{abstract}


\maketitle

\section{Introduction}

Starting with the pioneering observation of topologically protected chiral edge modes around a time-reversal-breaking two-dimensional photonic crystal~\cite{Haldane2008,Wang2009}, the last decade has witnessed the explosion of the field of {\em topological photonics}. Taking inspiration from condensed matter physics concepts such as topological insulators and quantum Hall effects, new exciting optical effects were found, which are paving the way to technological applications~\cite{Ozawa2018,Lu2014}.

So far, experiments have mostly addressed single-particle topological features, which are observable via the linear optical properties of the system: besides direct evidences of the topological order such as chiral edge states in different geometries, platforms, and spectral regions~\cite{Wang2009,Hafezi2013,Rechtsman2013,Jacqmin2014,Ningyuan2015,Lu2015}, remarkable results were the measurement of the band Berry curvature~\cite{Wimmer2017}, the observation of magnetic Landau levels~\cite{Schine2016}, of topological pumping~\cite{Kraus2012}, of anomalous Floquet edge states~\cite{Mukherjee2017,Maczewsky2017}, of synthetic dimensions~\cite{Lustig2018,Zilberberg2018,Dutt2019,Dutt2019a}.
Beyond linear optics, a great attention is nowadays devoted to the rich interplay between optical nonlinearities and topology: nonlinearity-driven topological phase transitions~\cite{Leykam2016} and self-localized states~\cite{Lumer2013} were anticipated for classical light, while the strongly correlated quantum Hall states of light predicted for ultra-strong nonlinearities~\cite{Cho2008,Umucalilar2012,Hafezi2013} are being actively investigated in circuit-QED systems~\cite{Roushan2017} and in atomic gases in a Rydberg-EIT configuration~\cite{Clark2019}.

One of the most promising applications of topological photonics concerns laser operation in topological systems displaying optical gain, the so-called {\em topological lasing}. As a first step, lasing into the zero-dimensional edge states of a one-dimensional Su-Schrieffer-Heeger (SSH) chain was proposed~\cite{Pilozzi2016,Solnyshkov2016} and experimentally demonstrated~\cite{St-Jean2017,Parto2018,Zhao2018}. Soon afterwards, lasing into the one-dimensional chiral edge states of a two-dimensional topological lattice was experimentally realized in suitably designed semiconductor laser devices~\cite{Bahari2017,Bandres2018}. Such topological lasers appear promising to solve a long-standing technological problem in opto-electronics, namely the realization of large-area devices offering high-power coherent emission~\cite{Longhi2018APL}: a pioneering theoretical work~\cite{Harari2018} has in fact pointed out that the topological protection against fabrication defects should make laser operation into topological edge states to remain single mode and to have a high slope efficiency even well above the laser threshold. 
This optimistic view was somehow questioned in~\cite{Longhi2018} for the specific case of semiconductor-based devices: using a standard model of laser operation in these systems, dynamical instabilities stemming from the combination of nonlinear frequency shifts and of the slow carrier relaxation time were predicted.

The purpose of this article is to build a generic theory of topological laser operation.
Going beyond the pioneering works~\cite{Harari2018,Longhi2018,Kartashov2019}, we identify a number of peculiar effects that directly stem from the chirality of the lasing mode and thus differentiate topological lasers from standard lasers. Keeping the complexity of the model at a minimum level, our attention will be focused on those general effects that play a central role in different realizations of topological laser devices. Such an analysis will provide a powerful conceptual framework in view of future studies of the complex nonlinear physics of specific realizations of topological laser devices and, on the longer run, will be a useful starting point to understand the fundamental quantum limits of topological laser operation. 

The structure of the article is the following. In Sec.~\ref{sec:model}, we review the basics of the Harper-Hofstadter model and we introduce our theoretical description of gain and losses. In Sec.~\ref{sec:WSG} we briefly review the chaotic behaviour in the presence of a spatially uniform gain. In Sec.~\ref{sec:WEG} we discuss how restricting gain to the edge of the lattice allows to obtain a single-mode laser emission that is robust against disorder. The peculiar features that stem from the chiral nature of the lasing mode are highlighted, as well as the limitations they are expected to impose on the laser performance. In Sec.~\ref{sec:PEG}, we investigate the effect of restricting gain to a finite strip of sites along one edge. For this geometry, the finite group velocity of the chiral edge mode turns out to be responsible for a marked distinction between convective and absolute instabilities, which is characterized in terms of noise-sustained structures and traveling wave amplification. Conclusions are finally drawn in Sec.~\ref{sec:conclusions}.

\section{The model}
\label{sec:model}

\begin{figure}[t!]
    \includegraphics[scale=0.5]{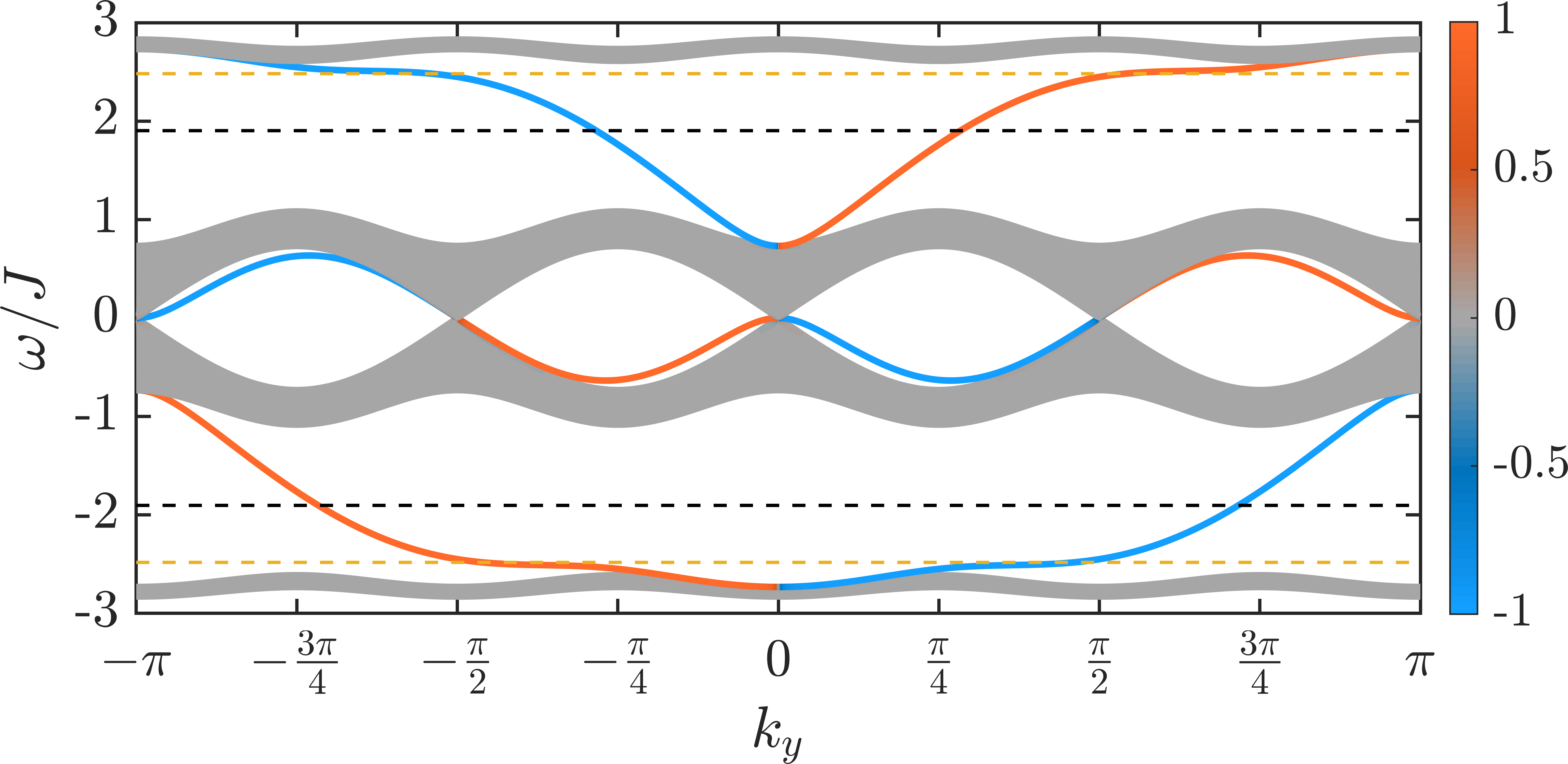}
	\caption{Energy bands of the conservative Harper-Hofstadter Hamiltonian \eqref{eq:consEOM} with flux $\vartheta = 1/4$ in a lattice of $N_x=399$ sites along $x$ and periodic boundary conditions along $y$. Blue vs. red color scale quantifies localization on the left or right edges. The horizontal black and orange lines indicate the WEG and PEG lasing frequencies shown in Fig.~\ref{fig:03}(c).}
	\label{fig:01}
\end{figure}

Since we are interested in the generic topological features, we will concentrate on an archetypal topological model, namely the bosonic Harper-Hofstadter (HH) model~\cite{Ozawa2018}. Modulo the extra pseudo-spin degree of freedom associated to the propagation direction around the ring cavities~\cite{Hafezi2011,Hafezi2013}, this model underlies the topological laser operation of~\cite{Bandres2018}. In the Landau gauge, the HH Hamiltonian reads~\cite{Hofstadter1976}:
\begin{align}
	H = \sum_{m,n} \Big\lbrace &\omega_0a_{m,n}^{\dagger}a_{m,n} 
	- J\big( a_{m,n}^{\dagger}a_{m+1,n} \nonumber \\
	&\qquad+ e^{-i 2\pi\vartheta m}a_{m,n}^{\dagger}a_{m,n+1} + \text{h.c.}  \big)
	\Big\rbrace,
	\label{eq:consEOM}
\end{align}
where the sum runs over all lattice sites, $\omega_0$ is the natural frequency of the microrings, and $a_{m,n}$ is the photon field amplitude at the site $(m,n)$. In the chosen gauge, the hopping amplitude along the $x$ direction is real and constant and equal to $J$, while hopping along $y$ involves an $x$-dependent phase. The strength of the synthetic magnetic field is quantified by the flux $\vartheta$ per plaquette in units of the magnetic flux quantum. For rational $\vartheta=p/q$, the bulk eigenstates distribute in $q$ bands characterized by non-vanishing topological Chern numbers. As a result, spatially finite lattices display chiral edge states unidirectionally propagating around the system and localized in the energy gaps between the bands. In what follows we will focus on the simple $\vartheta=1/4$ case, whose dispersion of band and edge states in a cylindrical geometry is sketched in Fig.~\ref{fig:01}(a). Since the lowest and highest bands have a non-vanishing Chern number $|C|=1$, for each of the main energy gaps a single edge state is present on each edge of the system.

Within the semiclassical theory of lasing for a broad-band gain medium~\cite{Scully1997}, losses and gain can be included as additional terms in the time-evolution of the classical field amplitudes $a_{m,n}$~\cite{Secli2017,Harari2018},
\begin{equation}
	\dot{a}_{m,n}(t) =-i\left[a_{m,n},H\right] 
	+ \left(\frac{P_{m,n}}{1+\beta|a_{m,n}|^2} - \gamma\right)a_{m,n},
	\label{eq:EOM}
\end{equation}
where the first term on the right-hand side gives the usual equations of the motion of the conservative HH model.
Here $\gamma$ accounts for the intrinsic resonator losses, $P_{m,n}$ determines the spatial profile of the gain, and $\beta$ sets the gain saturation level.
In our calculations, we start from an initial state with a small Gaussian noise and numerically simulate the evolution \eqref{eq:EOM} until its steady-state.

\section{Whole system gain (WSG)}
\label{sec:WSG}

\begin{figure}[t!]
    \mbox{
		{\hspace{-2pt}\includegraphics[scale=0.5]{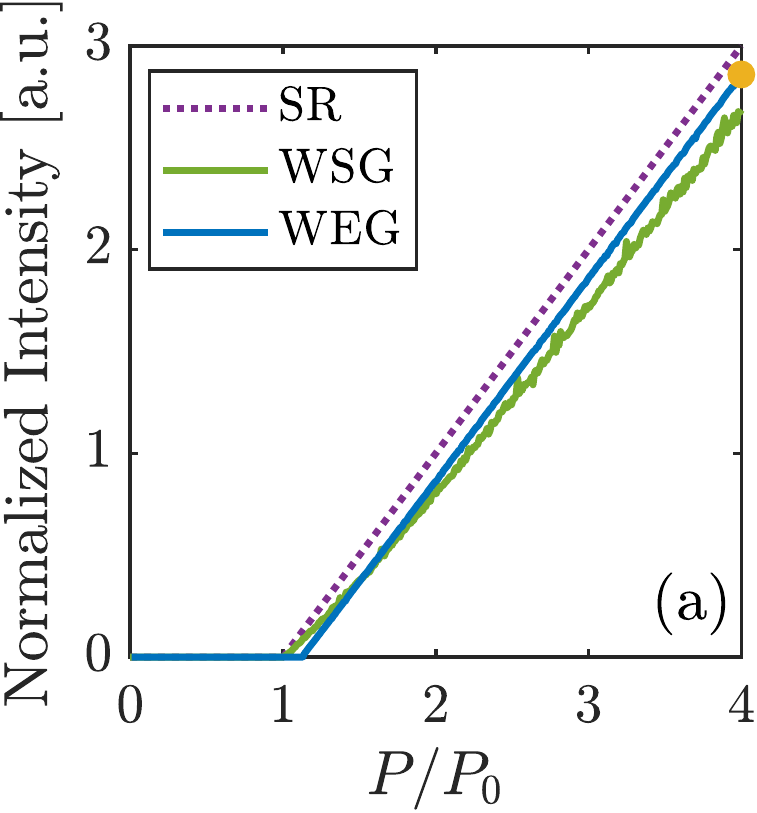}}
		{\hspace{4.0pt}\raisebox{2.55pt}{\includegraphics[scale=0.5]{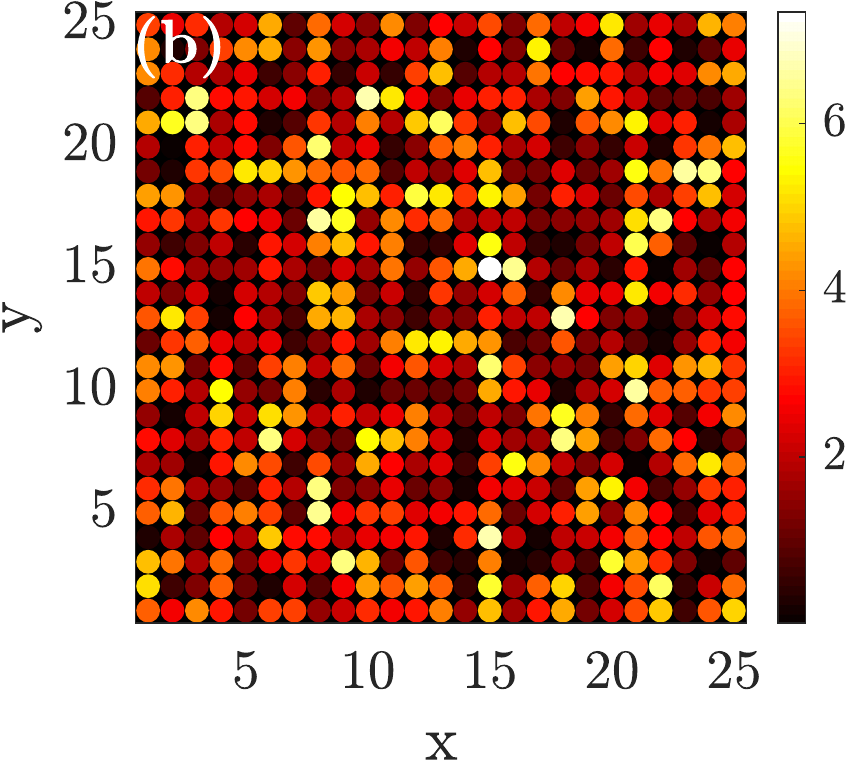}}}
	}
	\caption{Topological lasing in a $25 \times 25$ HH lattice with a flux $\vartheta = 1/4$  per plaquette. Left panel: total intensity $I_T=\sum_{m,n} |a_{m,n}|^2$ normalized to the number of amplifying sites vs. gain strength for different configurations: single resonator (SR), whole system gain (WSG) and whole edge gain (WEG).
	Right panel: snapshot of the typical intensity distribution at an arbitrarily chosen time $t=1000\gamma^{-1}$ in a WSG configuration. If not differently specified, we have taken $\beta=1$.}
	\label{fig:02}
\end{figure}

We start our discussion by reviewing the case of a  spatially uniform $P_{m,n}=P$ gain.
Fig.~\ref{fig:02}(a) shows how the lasing threshold remains very close to the single-resonator (SR) value $P_0\doteqdot \gamma$ analytically extracted from \eqref{eq:EOM}, the slope efficiency  $dI_T/dP$ is only slightly lower than the single-resonator value, and the laser emission is spread throughout the whole system. However, due to complex mode-competition effects, the intensity distribution is very inhomogeneous in space [Fig.~\ref{fig:02}(b)] and no monochromatically oscillating steady state is ever reached. This strong spatio-temporal modulation persists indefinitely (see Supplemental Video 1) and is due to the simultaneous lasing into many modes that interfere and interact with each other via the intrinsic nonlinearity of the model. Such chaotic behaviours are very common in laser arrays unless some specific stabilization scheme is introduced~\cite{Winful1988,Hohl1997,Katz1983,Longhi2018APL}. As one can see in the Supplemental Video 1, while the chaotic dynamics of the bulk does not appear to display any specific signature of the non-trivial topology, the intensity distribution on the edge keeps circulating around the system.

\section{Whole edge gain (WEG)}
\label{sec:WEG}

\begin{figure}[t!]
   	\mbox{
		{\hspace{-2pt}\raisebox{1.8pt}{\includegraphics[scale=0.5]{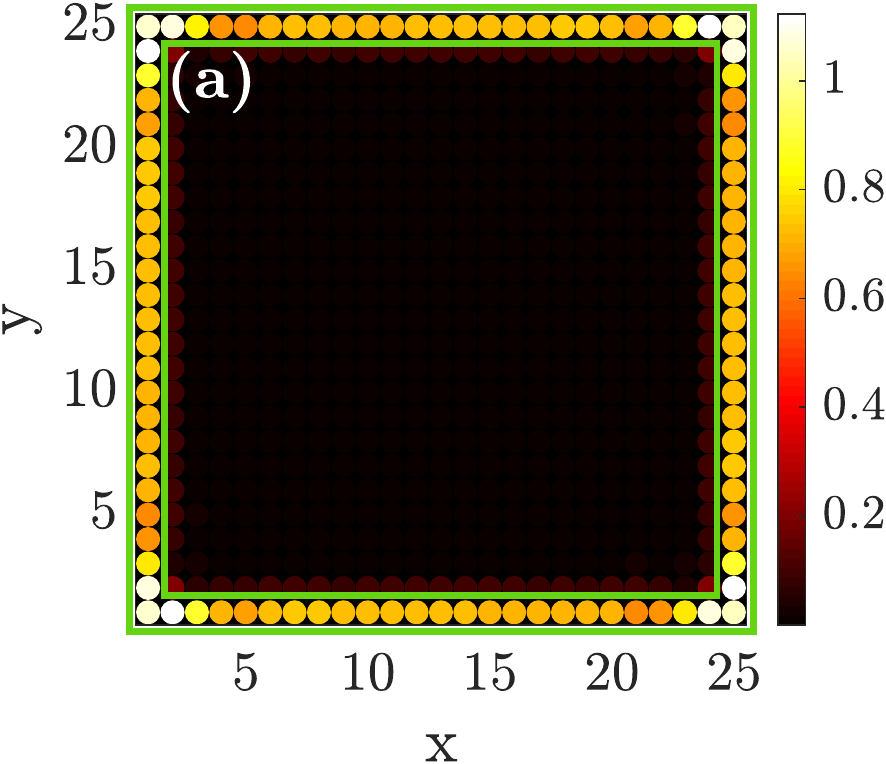}}}
		\includegraphics[scale=0.5]{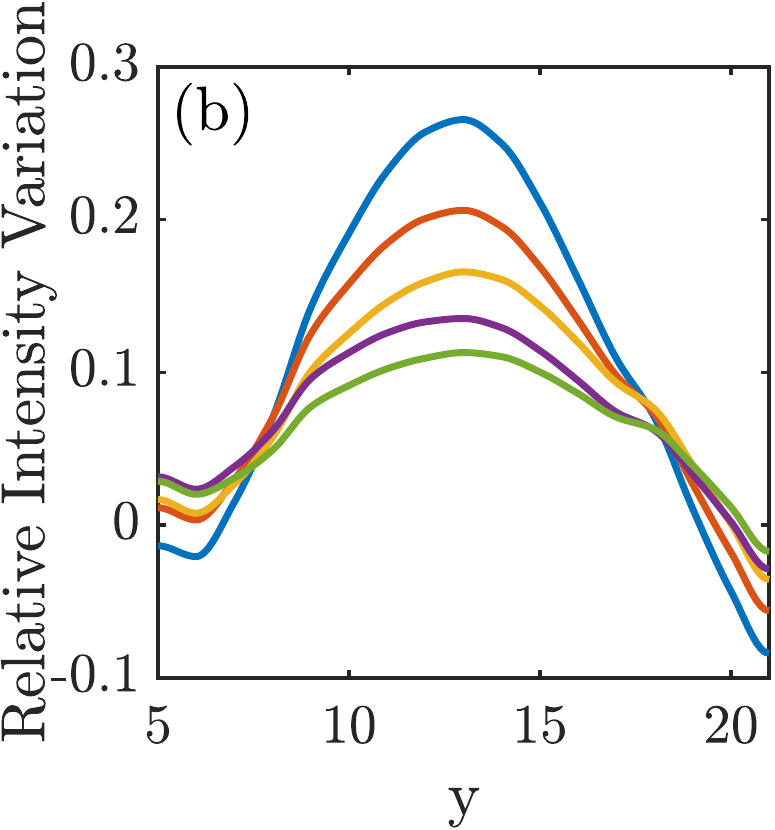}
	}\\
	\smallskip
	\mbox{
		\hspace{-6pt}\includegraphics[scale=0.5]{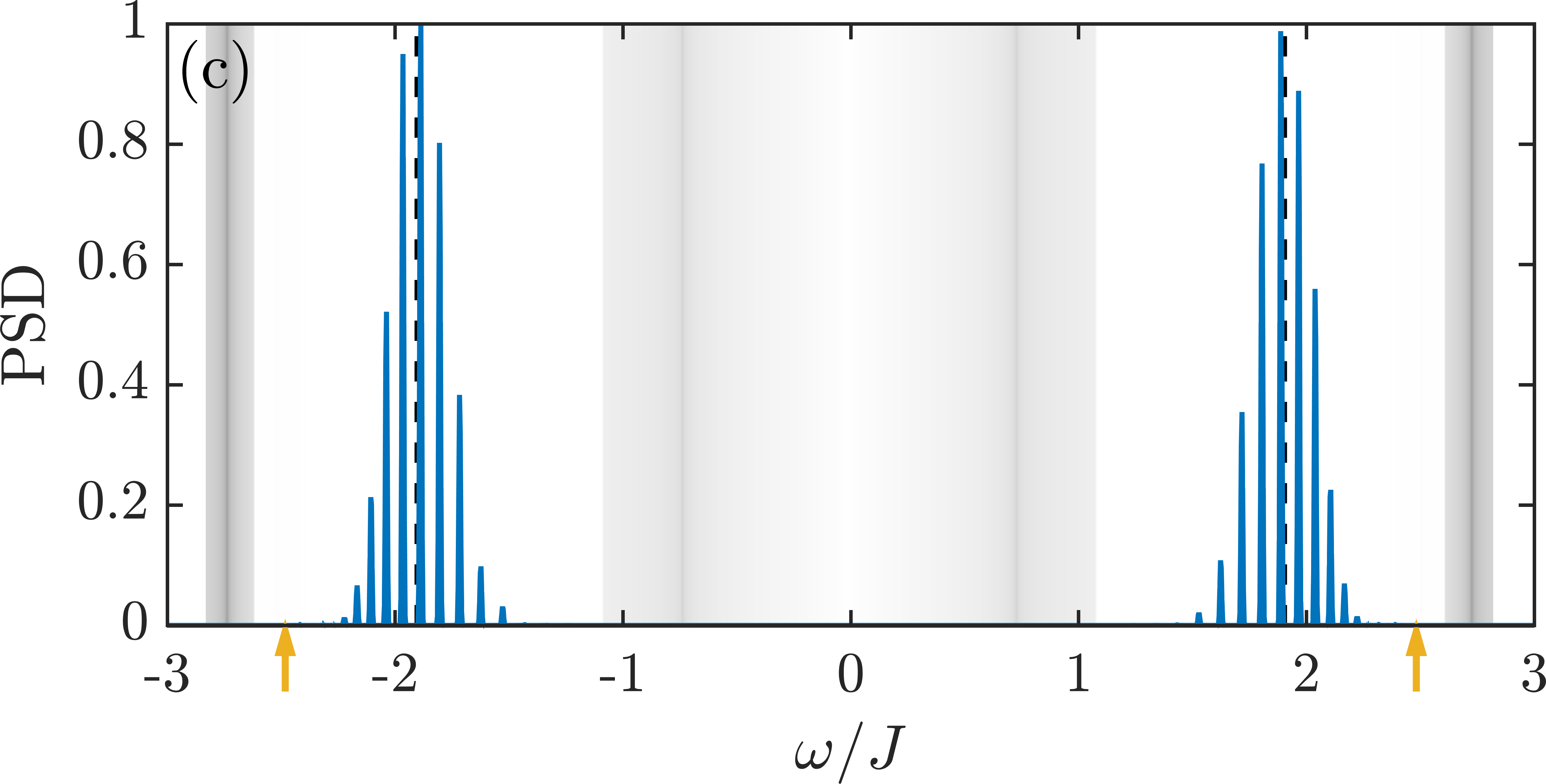}
	}
	\caption{
	Topological lasing in a $25 \times 25$ HH lattice with $\vartheta = 1/4$ in a one-site-thick WEG configuration. Panel (a): snapshot of the steady-state intensity distribution. The green rectangle indicates the amplifying sites.
	Panel (b): cuts of the intensity distribution along the $x=1$ line at times (from top to bottom) $\gamma t=43.85, 51.65, 59.45, 67.20, 75.00$. Panel (c): normalized realization-averaged power spectral density (PSD). The dashed lines indicate the center of mass of the distribution. For comparison, the orange arrows indicate the lasing frequency for  a $1\times 15$ PEG. The gray shading indicates the density of states of the bands in Fig.~\ref{fig:01}(a).}
	\label{fig:03}
\end{figure}

\begin{figure*}[t!]
    \centering
    \mbox{
        \includegraphics[scale=0.5,valign=t]{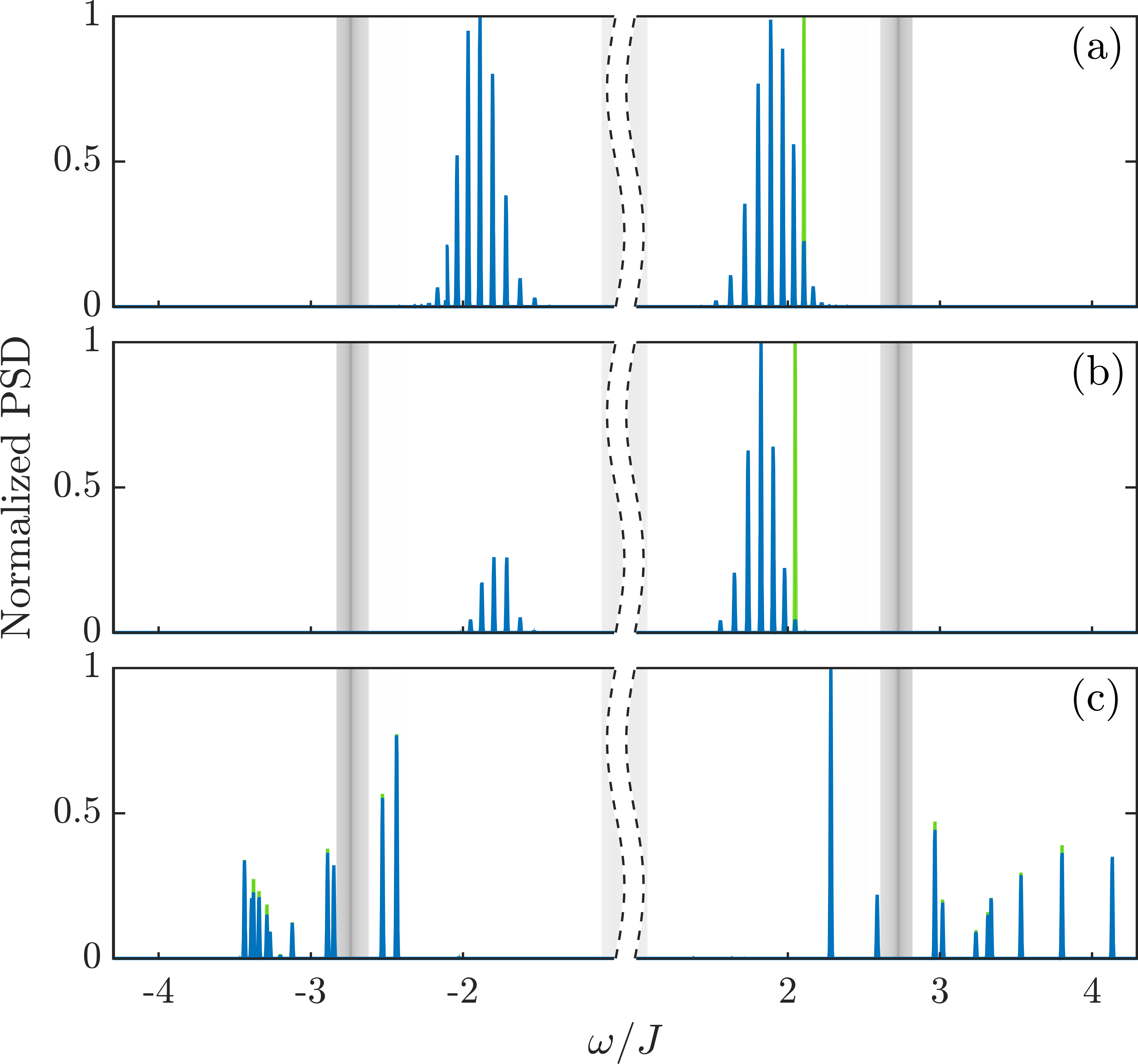}
        \hspace{18pt}
        \raisebox{-1.5pt}{\includegraphics[scale=0.5,valign=t]{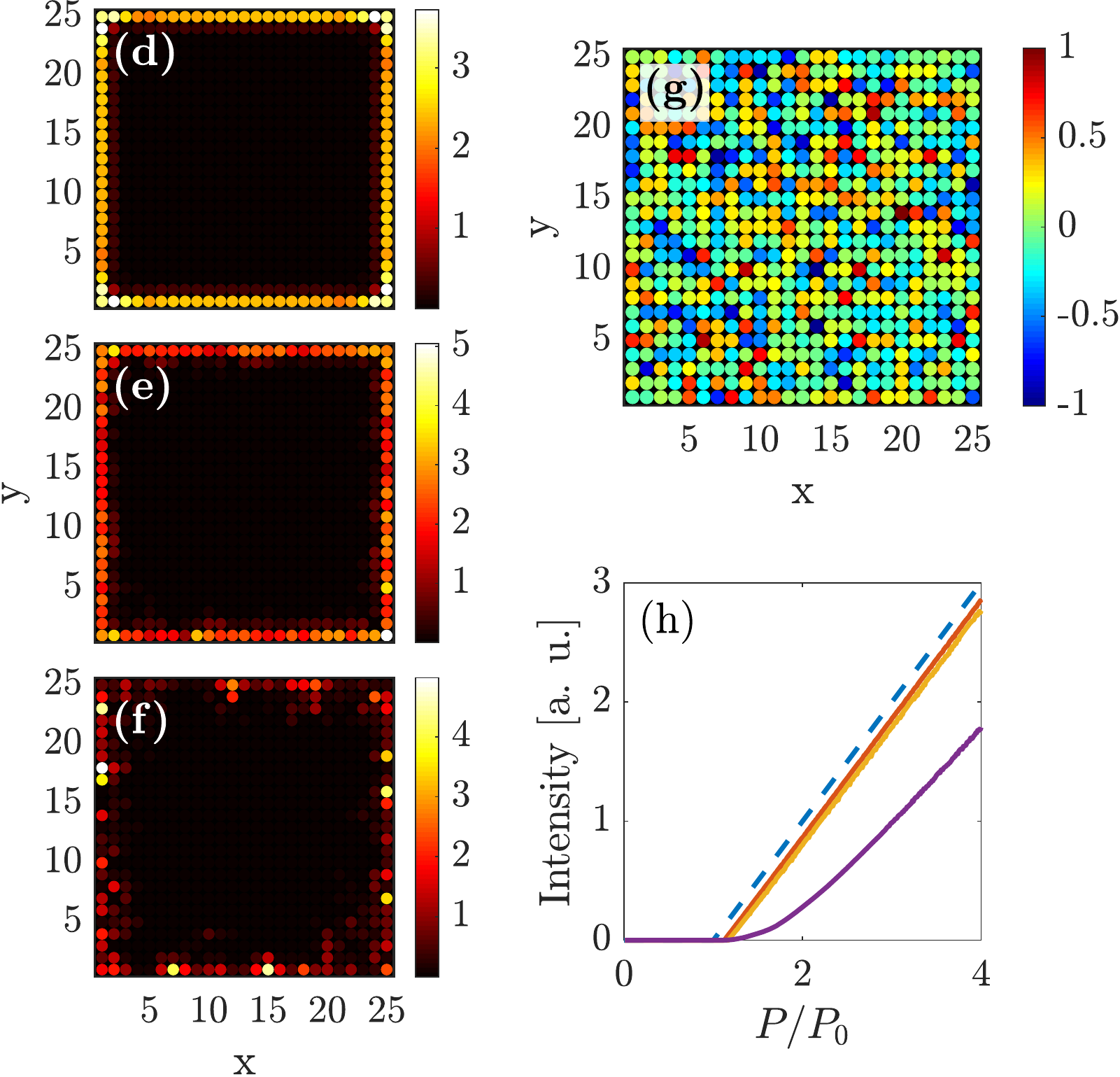}}
    }
	\caption{Panels (a)-(c): normalized spatially integrated power spectral density (PSD) in the WEG configuration without disorder (panel (a)) and with a disorder strength equal to $\sigma(U)/J = 0.4$ (panel (b)) and to $\sigma(U)/J = 1.2$ (panel (c)). The PSD for a single random realization of lasing is displayed in green, while the average over multiple realizations [5000 in panel (a), 2500 in panels (b)-(c)] is shown in blue. The shaded areas indicate the density of states of the bands in the absence of disorder and the central band region has been cut out for visualization convenience. Panels (d)-(f):  snapshots of the typical intensity distribution at an arbitrarily chosen late time $t = 500\gamma^{-1}$ for the same values of disorder as in (a)-(c). Panel (g): realization of the disorder used (upon rescaling) for all other simulations in the figure. Colors indicate the frequency shift of the different sites for a disorder strength $\sigma(U)/J = 0.1$. Panel (h): normalized emitted intensity as a function of gain strength for a single resonator (blue dashed line), for the non-disordered case (solid red line) and for a few disordered cases with $\sigma(U)/J=0.4$ (solid yellow line) and $\sigma(U)/J=0.4$ (violet line).}
	\label{fig:04}
\end{figure*}

A natural strategy to favor laser emission in the topological edge states is to restrict the gain to the sites on the geometrical border of the system, as experimentally implemented in~\cite{Bandres2018}~\footnote{Note that topological lasing in~\cite{Bahari2017} was operated under a WSG. The physical reason why bulk mode lasing was suppressed in this experiment is presently under investigation.}. Fig.~\ref{fig:03}(a) recovers the predictions of earlier theoretical work~\cite{Harari2018} and displays a stable monochromatic single mode oscillation in a topological edge mode of the system~\footnote{Note that the dynamical instabilities anticipated in~\cite{Longhi2018} were due to specific features of semiconductor lasers, in particular to the presence of a slow carrier reservoir that induces site-dependent nonlinear frequency shifts. They are absent in our simple model of lasing. A theoretical study of the stability of edge state lasing will be the subject of a future work.}. The slope efficiency (i.e. the slope of the blue curve in Fig.~\ref{fig:02}(a) right above the threshold) is very close to the single-site one and the slightly increased threshold $\tilde{P}_0\gtrsim P_0$ is due to the weak but finite penetration of the edge mode into the non-amplifying bulk sites. 
Given the broadband gain used in the calculations, the oscillation frequency occurs with the same probability in either gap of the band structure [Fig.~\ref{fig:03}(c)]: as expected from the band structure shown in Fig.~\ref{fig:01}, the lasing mode will have opposite chirality depending on which edge mode is selected.  

\subsection{Consequences of the chirality of the lasing mode}

This general picture of topological lasing~\cite{Harari2018,Longhi2018APL} is the starting point to investigate the subtle physical consequences of the chirality of the lasing modes that are the core subject of this article. 

As a first result, Fig.~\ref{fig:03}(c) shows that the lasing frequency is randomly chosen among a number of available modes located around the gap centers. Since the penetration of the edge mode in the bulk is minimum at the center of the energy gap, lasing will preferentially occur in this frequency region that maximises the overlap with the amplifying sites and thus the effective gain. 
As it happens in ring lasers, edge modes are discretized according to a round-trip quantization condition around the perimeter of the system. This gives a frequency spacing $\Delta\omega\simeq 2\pi\,v_g/L$ where $v_g$ is the edge mode group velocity and $L$ is the perimeter. The approximately equal spacing of the modes is due to the weak curvature of the edge mode band that is visible in Fig.~\ref{fig:01}.
Even though the mode spacing can be very small in large lattices, once a lasing mode has been selected, the single-mode emission remains stable for indefinite times in the absence of noise. The overall width of the distribution is determined by the $k$-dependent spatial overlap of edge modes with the gain region, which introduces an effective frequency dependence of the gain.

As an even more remarkable feature, Fig.~\ref{fig:03}(b) displays a series of longitudinal cuts of the intensity profile along the $x=1$ left edge for different times separated by an (approximate) round-trip time $T_{\mathrm{rt}}=L/v_g$. The intensity modulation due to the initially noisy state relaxes away on a much slower time-scale than all other microscopic scales, including $T_{\mathrm{rt}}$. As an illustrative example, Supplemental Video 2 shows an intensity bump traveling in the clockwise direction around the system and slowly fading away. This ultra-slow relaxation rate is a consequence of the Goldstone theorem which imposes (at least) a $k^2$ behaviour for the imaginary part of the complex frequency of the long-wavelength collective modes corresponding to spatially slow fluctuations of the laser emission~\cite{Wouters2007,Loirette-Pelous2019}.

\subsection{Robustness to disorder}

To complete the picture, it is important to briefly investigate the robustness of these features against static disorder. Some first remarks on the effect of disorder were reported in~\cite{Harari2018}.

The most straightforward way of including disorder in our model is to introduce a random frequency shift of the natural frequencies of the cavities.
In Fig.~\ref{fig:04} we take the on-site disorder $U$ to have a Gaussian distribution with standard deviation $\sigma(U)$. A specific realization of disorder is displayed in panel (g) for the $\sigma(U)/J = 0.1$ case. The disorder used in the cases $\sigma(U)/J=0.4,\,1.2$ is obtained by simply rescaling this distribution. In addition to this ``non-magnetic'' disorder that is common to all systems, note that microrings-based implementations like the one in~\cite{Bandres2018} can also host another source of disorder, called ``magnetic'' disorder since it couples the two pseudo-spin states~\cite{Hafezi2011,Harari2018}. A study of this latter disorder goes beyond the scope of our work.

Thanks to the topological protection of the edge mode and its ability to circumnavigate impurities and defects, the intensity distribution for a WEG configuration remains spatially localized on the edge up to large values of the disorder strength comparable to the bandgap [panels (d)-(f)]. As one can see by comparing the different curves in panel (h), moderate values of disorder only slightly increase the lasing threshold, while the slope efficiency is almost unaffected. The unidirectional chiral motion of the lasing edge mode guarantees an efficient phase locking of the emission at different points along the edge and the laser operation remains firmly single mode [green lines in panels (a)-(b)]. 

Only when the disorder gets comparable to the energy band gap, the laser emission breaks into several independently lasing regions and the emission acquires a multi-mode and multi-frequency character, as shown in the intensity distribution in panel (f) and in the spectrum in panel (c). Correspondingly, one can see in panel (h) that the sharp threshold transforms into a smooth, progressive switch-on. As compared to the WSG case, the spatial separation of the different lasing modes makes the temporal fluctuations of the intensity profile less apparent than in Fig.~\ref{fig:02} and Supplemental Movie 1.

The blue lines in panels (a)-(c) show a statistical analysis of the emission frequency over many realization of laser operation with the same realization of the Gaussian disorder \footnote{The study of a single disordered realization is physically more meaningful, in this case, than averaging over multiple disorder realizations; it models a specific sample that has a single and immutable disordered configuration.}.
As long as the disorder is moderate and the lasing mode keeps extending around the whole system, the discretization of the modes is preserved [blue lines in panels (a-b)]. 
For stronger disorder, when many modes are simultaneously and independently lasing, the emission spectrum for a single realization matches the averaged one, so the distinction between the green and the blue curve is no longer visible in panel (c).

\section{Partial edge gain (PEG)}
\label{sec:PEG}

\begin{figure}[t!]
    \mbox{
	    \raisebox{0.1pt}{\includegraphics[scale=0.5]{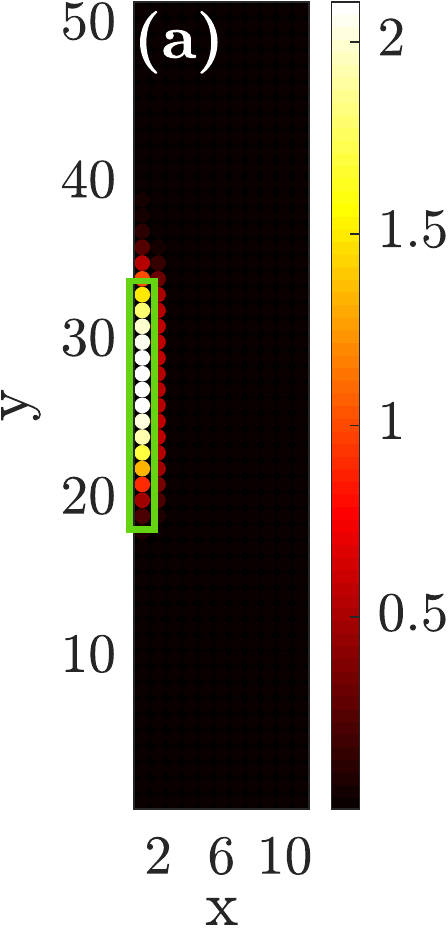}}
        \hspace{9pt}
	    \includegraphics[scale=0.5]{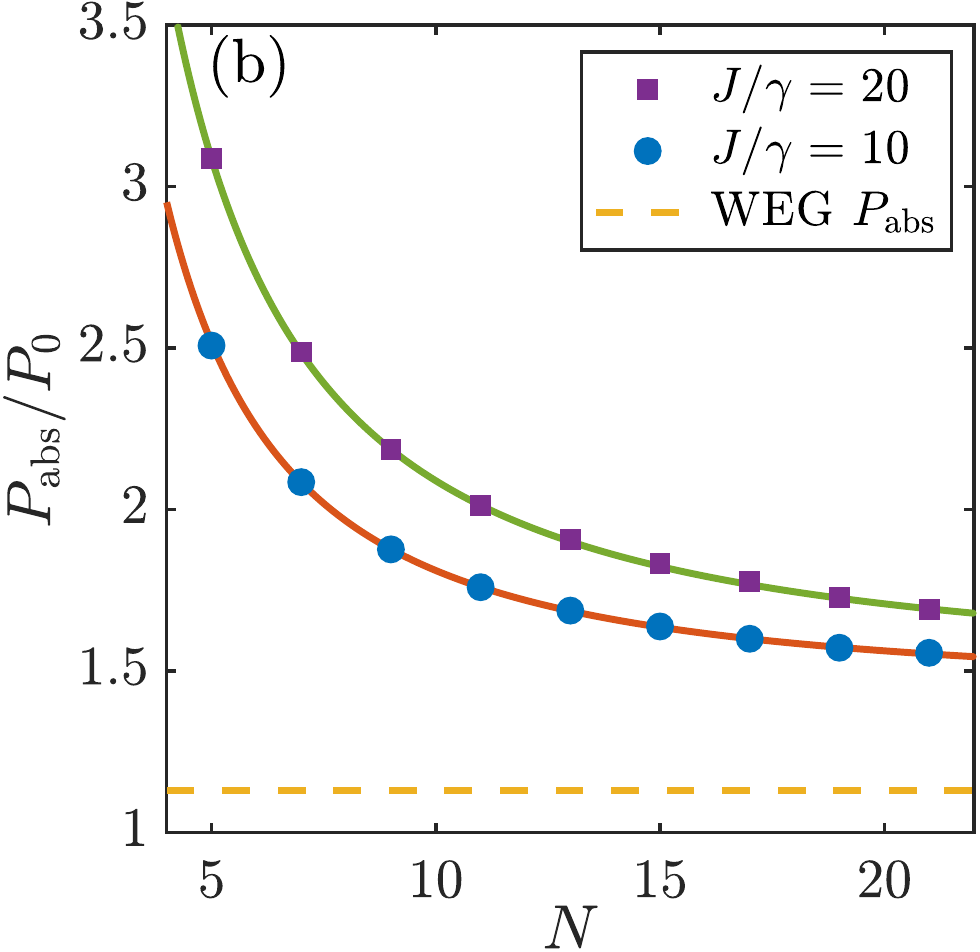}
	}\\
	\smallskip
	\includegraphics[scale=0.5]{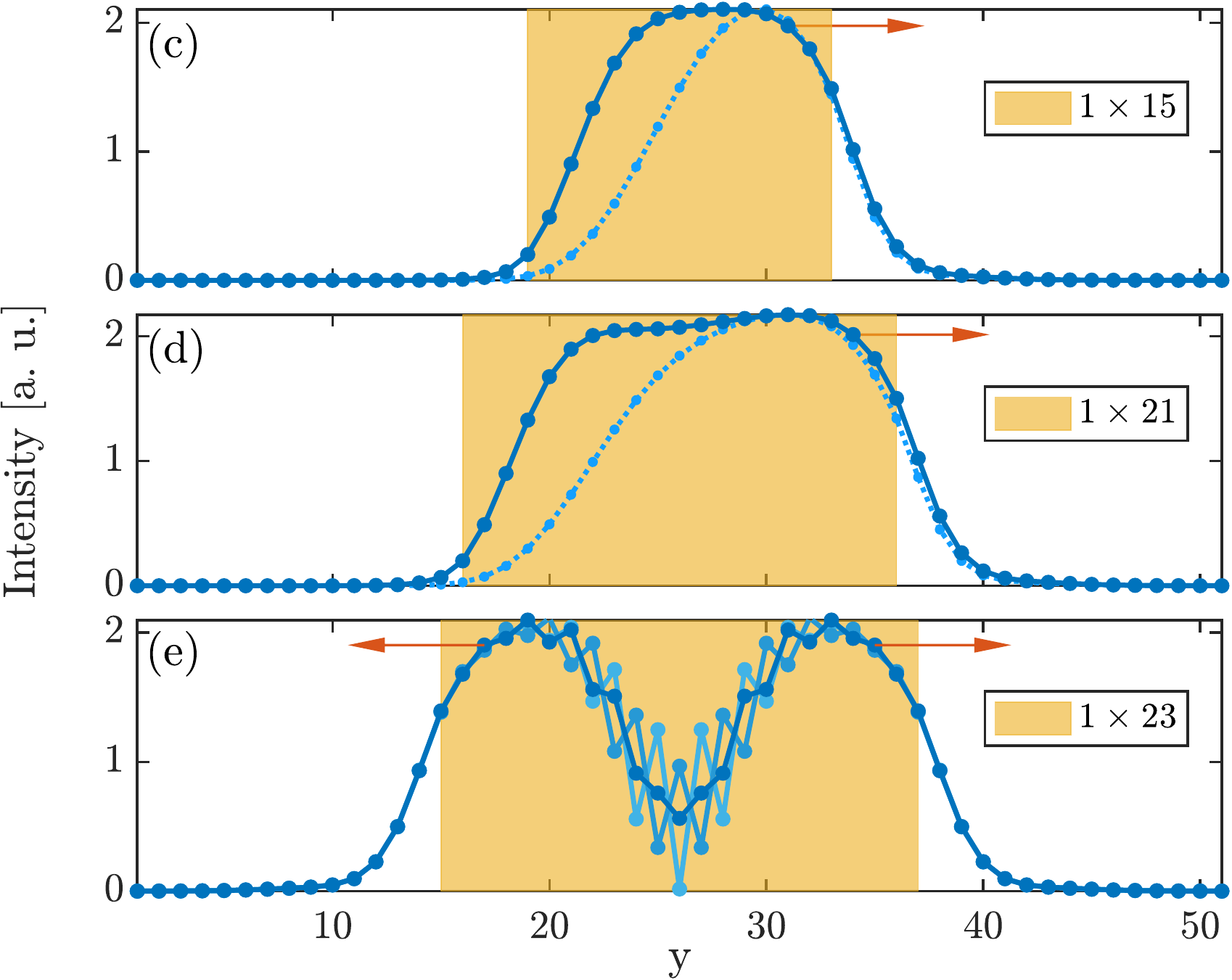}
	\caption{Panel (a): steady-state intensity distribution for a $1\times 15$ PEG in a large $11 \times 51$ lattice. The green rectangle indicates the amplifying sites. Panel (b): lasing threshold for $1\times N$ PEG with different $J/\gamma=20, 10$ (solid lines). Lasing threshold for a one-site-thick WEG case (dashed line). Panels (c)-(e): cuts of the intensity distribution along the $x=1$ line for different PEG geometries (see legends). The shaded area indicates the amplifying sites. The different curves in (c)-(d) refer to the steady-state for different gain strengths $P/P_0 = 4$ (solid blue), $P/P_0 = 2$ (dotted, light blue; to facilitate reading, these curves have been rescaled to have the same maximum as the solid blue ones); the different curves in (e) refer to different times separated by $0.05\gamma^{-1}$.}
	\label{fig:05}
\end{figure}

\begin{figure*}[t!]
	\centering
	\includegraphics[scale=0.5]{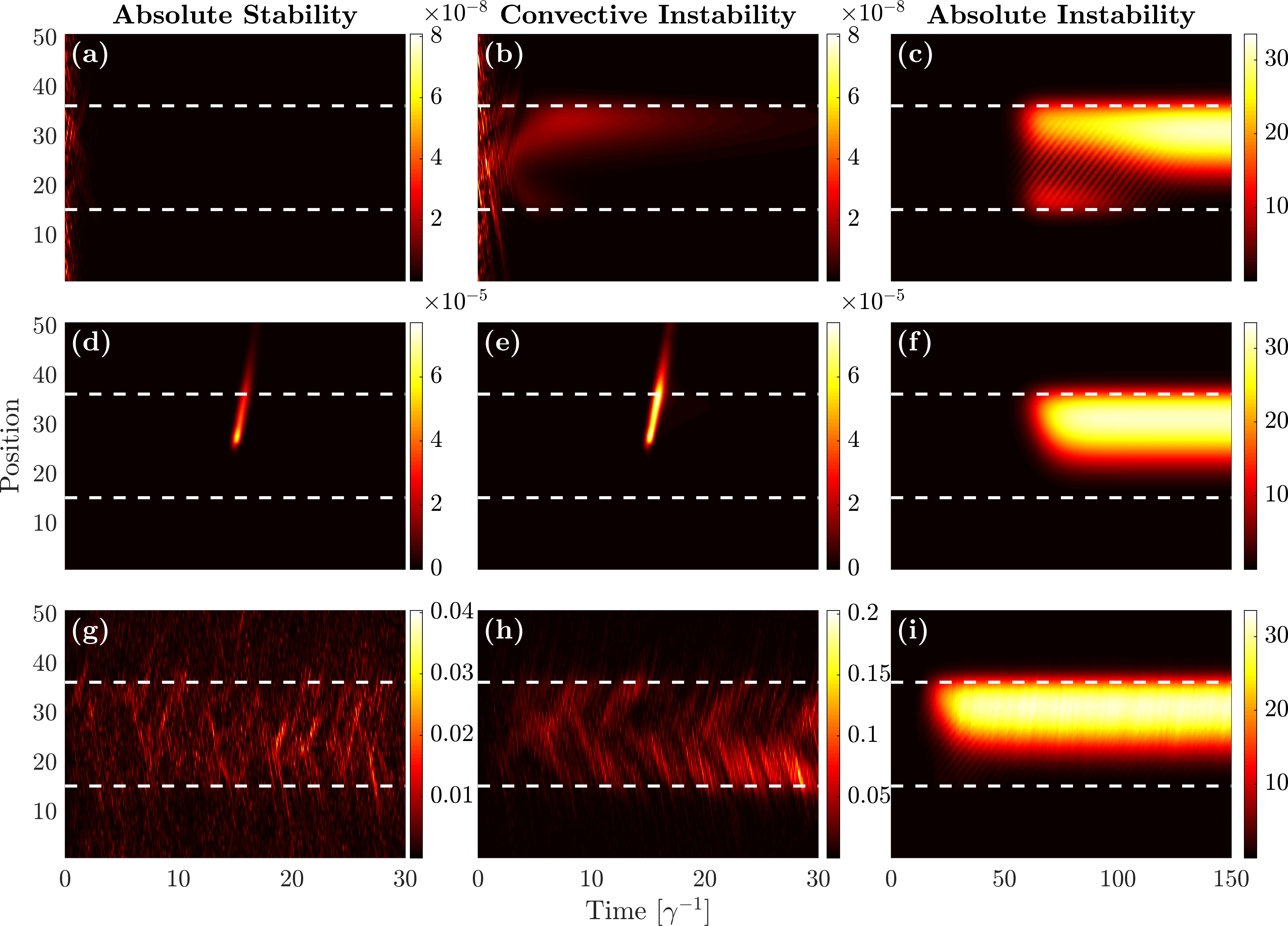}
	\caption{Spatio-temporal intensity patterns on the left border of a $11 \times 51$ lattice with a $1 \times 21$ amplified strip located at the center of the left side and indicated in all panels by the dashed white lines.
	In the first row there is a noisy seed only at $t=0$, while on the third one there is noise at all times. In the second row there is no noise, but there is a Gaussian pulse at frequency $\omega/J = 1.9$ localized on the central site and centered in time at $t=15$, with standard deviation $\sigma_t = 0.3$. All the times are measured in units of $\gamma^{-1}$. For homogeneity, in the AI case we restricted to realization of lasing into the clock-wise propagating topological mode localized on the left edge. In this figure,  $\beta=0.01$ was taken.}
	\label{fig:06}
\end{figure*}

Since the ultra-slow relaxation of long-wavelength fluctuations discussed in the previous Section is likely to compromise the coherence of the emission against quantum noise~\cite{Amelio2019}, it is interesting to explore a configuration where gain is restricted to a $1\times N$ finite strip of sites along an edge. A related geometry was experimentally considered in~\cite{Bandres2018}.

In this case, a dramatically faster relaxation can be anticipated since any perturbation is rapidly expelled by the chiral motion into the surrounding non-amplified edge region. Furthermore, while in the WEG configuration the round-trip quantization around the system perimeter gives a topologically protected winding number~\cite{Mermin1979} characterizing the lasing mode as in standard ring lasers, in the present PEG configuration the lasing region is an open segment, for which no topologically protected winding number exists; as a result, the spatial profile of the lasing mode is able to continuously relax towards its optimal shape. 

\subsection{Spatial structure of the lasing mode}

This expected behaviour is confirmed in Fig.~\ref{fig:05}. A steady-state with a stable monochromatic oscillation is indeed quickly reached on a microscopic time-scale. 
For moderate values of $N$ [panels (c) and (d)], all the emission is efficiently funneled into one of the two modes with opposite chiralities, randomly chosen at each realization. Given their relatively large frequency-separation of order $J$, one can anticipate that in practice one of them will be privileged by the small frequency-dependence of the gain~\footnote{Note that the pseudo-spin degree of freedom  in~\cite{Bandres2018} allows for more complex field configurations where modes of both chiralities are excited even in a monochromatic steady-state. As discussed there, more complex ring-resonators are then required to select a specific chirality.}.

The selected chirality reflects in the spatial asymmetry of the intensity profile within the amplifying region. This asymmetry is clearly visible on the dotted light-blue lines in panels (c) and (d) as a growing intensity along the positive-$y$ chiral propagation direction. This asymmetry is still visible but less marked on the solid blue lines calculated for a higher gain far above the threshold, for which the light intensity displays within the amplifying region a faster growth towards the saturated value.
Irrespectively of the gain strength, the chirality of the lasing edge mode is also apparent in the significant amount of light emission from the non-amplifying edge sites located just downstream of the amplifying region, while the ones located in upstream direction remain dark. In Fig.~\ref{fig:05}(c)-(d), this corresponds to a much more pronounced tail of the intensity distribution on the right-hand side of the amplifying region marked in yellow.

The situation is very different for large values of $N$. In this case, mode competition is not able to isolate a single mode and lasing simultaneously occurs in modes of both chiralities, [panel (e)]. Nonetheless, local gain saturation effects are still able to keep the two chiralities almost spatially separated with a net outward flow (red arrows). The fringes that are visible in the central region arise from interference of the two lasing modes and oscillate at their frequency separation of the order of $J$.

\subsection{Convective vs.~absolute instability}

Additional intriguing features of the PEG case are found in the dependence of the lasing threshold on the strip length $N$ plotted in Fig.~\ref{fig:05}(b). As expected the threshold decreases for growing $N$, but a numerical fit of the form $a N^{-b}+c$ (solid lines) clearly shows that the large-$N$ limit remains significantly higher than the WEG threshold (dashed line).

An explanation for this remarkable finding is offered by the distinction between convective and absolute instabilities, a well-known phenomenon in the theory of nonlinear dynamical systems and in hydrodynamics~\cite{Cross1993,Deissler1985}. The {\em absolute instability} (AI) corresponds to the standard dynamical instability of the zero-field state above a threshold $P_{\mathrm{abs}}$. The {\em convective instability} (CI) is instead a weaker form of instability that is found whenever the exponential growth of a perturbation for $P>\tilde{P}_0$ is overcompensated by its quick motion at $v_g$: in this CI regime, even though the {\em peak} amplitude of the moving perturbation grows in time, its {\em local} value at any given spatial location quickly decreases back to zero. When the amplifying region is spatially finite as in our PEG case, any perturbation immediately disappears upon entering the external lossy region. This distinction between CI and AI explains why the laser instability is only observed above the higher AI threshold $P_{\mathrm{abs}}>\tilde{P}_0\gtrsim P_0$. This phenomenon cannot occur in the WEG case where the closed shape of the amplifying region does not allow the perturbation to escape from it~\footnote{The transition between WEG to PEG occurs when the length of the non-pumped interval largely exceeds the absorption length along a (non-pumped) edge.}.

Further evidences of the role of the convective instability in the PEG configuration are offered by the dependence of the lasing operation on the group velocity $v_g$. 
As we have seen in the previous section, a lasing frequency next to the gap centers [Fig.~\ref{fig:03}(c)] is chosen in the WEG case so to maximize the spatial overlap with gain. In the PEG case, instead, the location of the absolute threshold $P_{\mathrm{abs}}$ is dominantly controlled by $v_g$, so the AI is first reached by edge modes located next to the outer edge of the gaps (orange arrows) for which $v_g$ is lower. 
A more subtle feature is visible in Fig.~\ref{fig:05}(b). On one hand, the WEG threshold (dashed line) stays constant at $\tilde{P}_0\gtrsim P_0$ when $J/\gamma$ (and thus $v_g$) is increased. On the other hand, the PEG threshold at $P_{\mathrm{abs}}$ monotonically grows when $J$ and consequently $v_g$ are increased (squares vs. circles). 

\subsection{Noise-sustained structures}

\begin{figure*}[t!]
	\centering
	\mbox{
    	\adjustbox{valign=c}{
            \includegraphics[valign=t,scale=0.5]{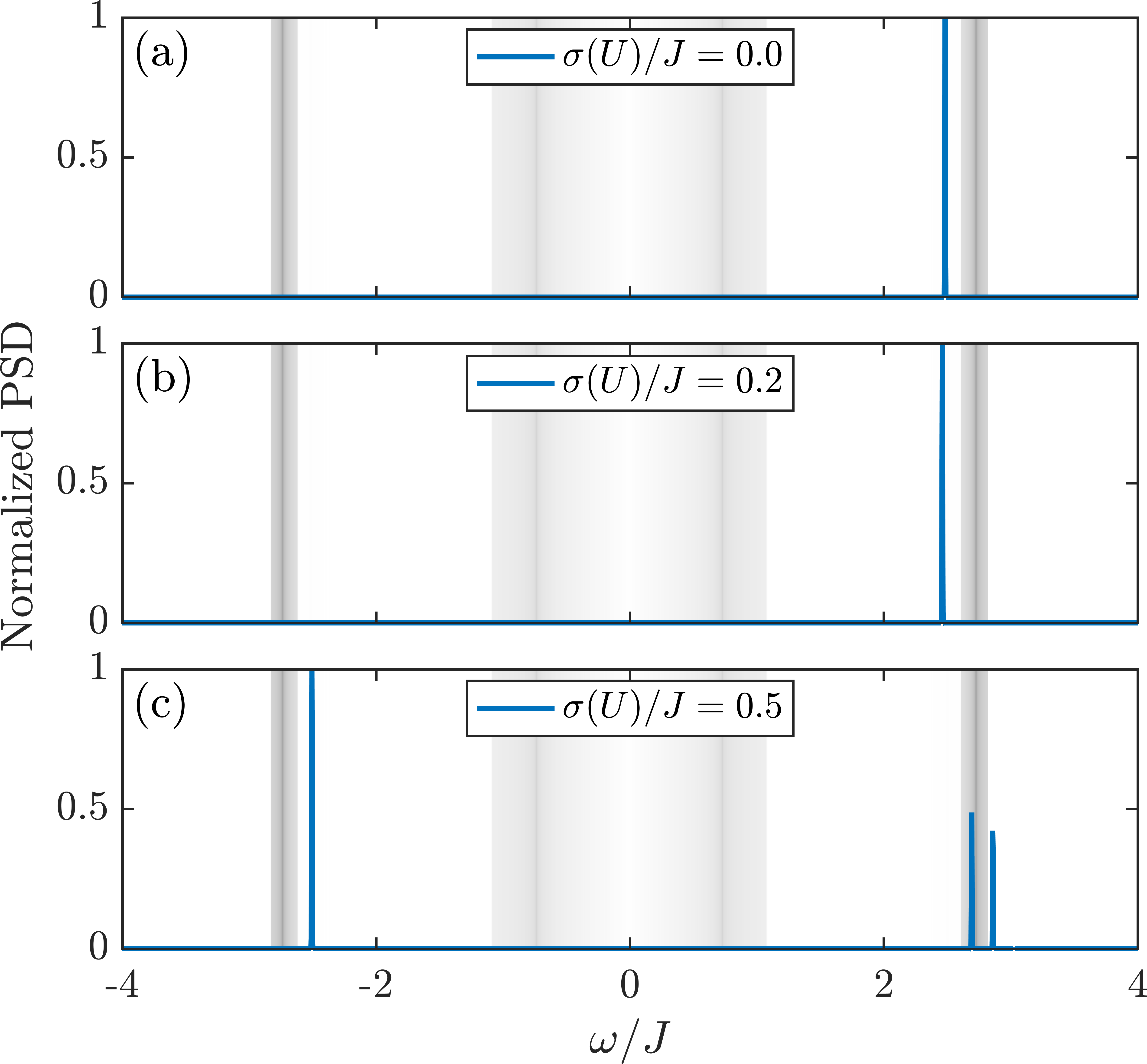}
        }
    	$\;\;$
        \adjustbox{valign=c}{
    	    \includegraphics[valign=t,scale=0.5]{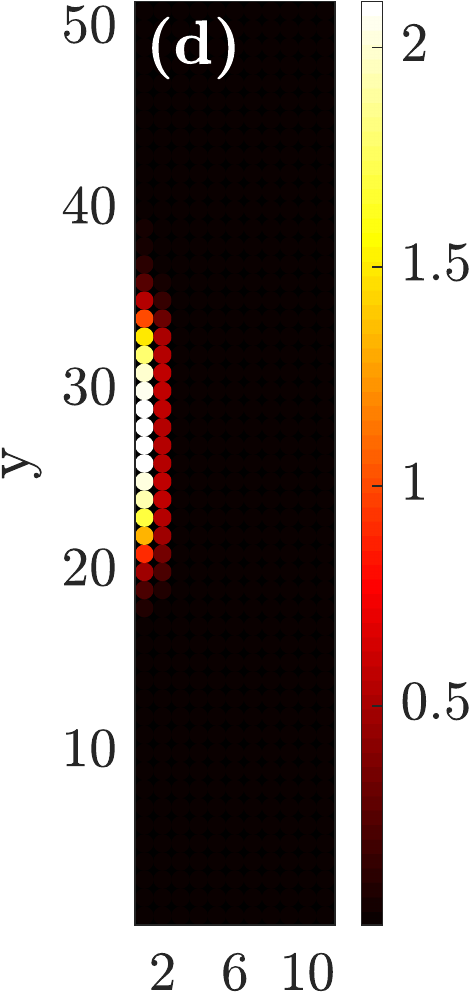}
    	    $\,$
    	    \includegraphics[valign=t,scale=0.5]{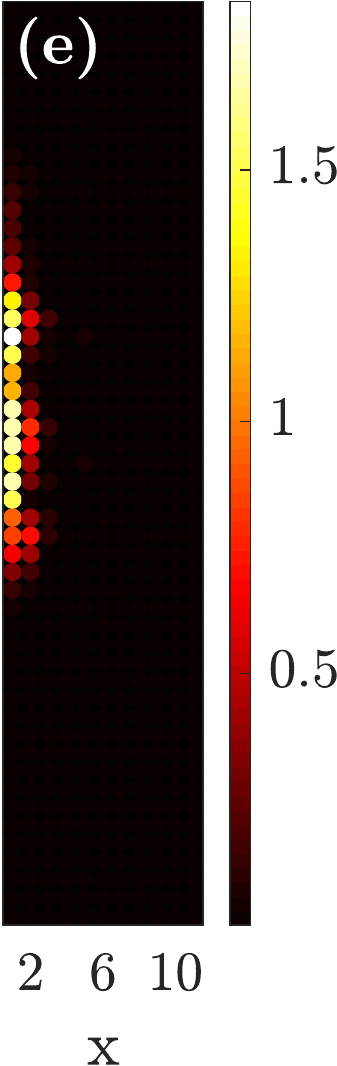}
    	    $\,$
    	    \includegraphics[valign=t,scale=0.5]{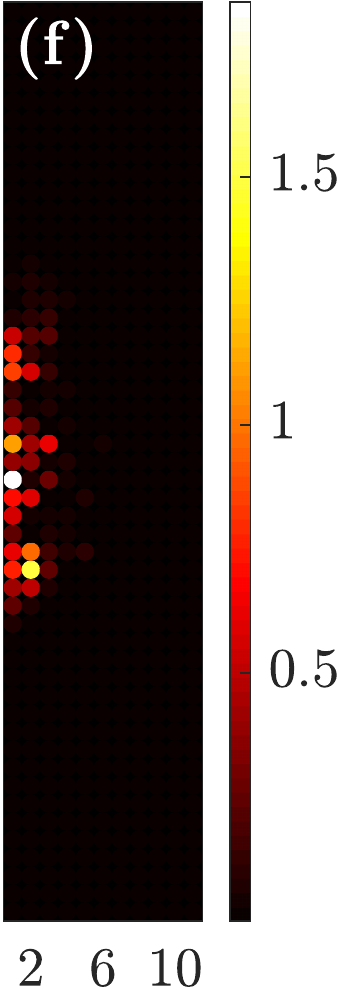}
    	    $\;\;$
    	    \includegraphics[valign=t,scale=0.5]{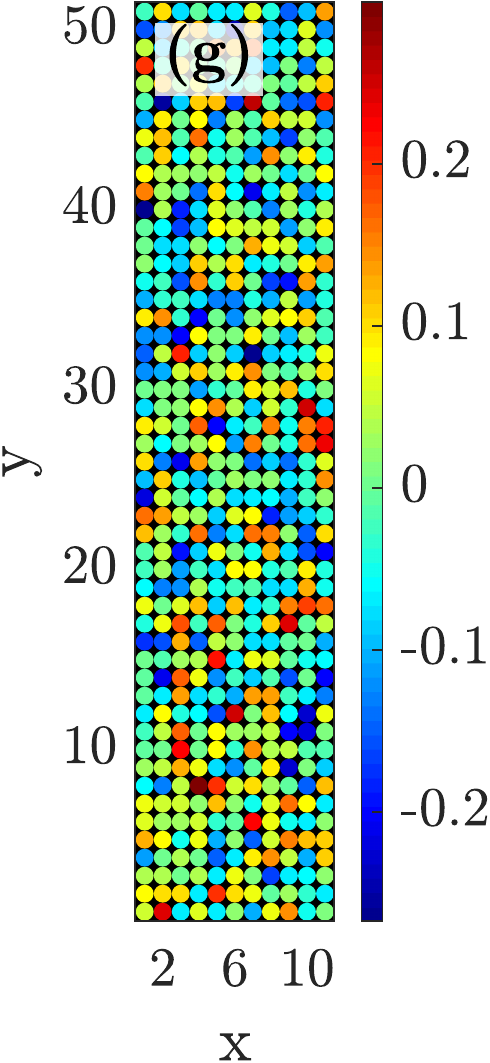}
    	}
	}
	\caption{Panels (a)-(c): normalized spatially integrated power spectral densities for different values of the disorder strength $\sigma(U)/J = 0$ [panel (a)], $0.2$ [panel (b)], $0.5$ [panel (c)]; the shaded areas indicate the density of states of the bands in the absence of disorder. Panels (d)-(f): snapshots of the intensity distribution at a late time $t = 500\gamma^{-1}$ for the same configurations of panels (a-c). Panel (g): frequency shift of the different sites for a disorder strength $\sigma(U)/J = 0.1$; upon a suitable rescaling, this realization of the disorder is used in all panels. Same geometry with a $1\times 15$ amplifying strip in a $11\times 51$ lattice as in Fig.~\ref{fig:05}(a). 
	}
	\label{fig:07}
\end{figure*}

\begin{figure}[t!]
	\centering
	\mbox{
		\includegraphics[valign=t,scale=0.5]{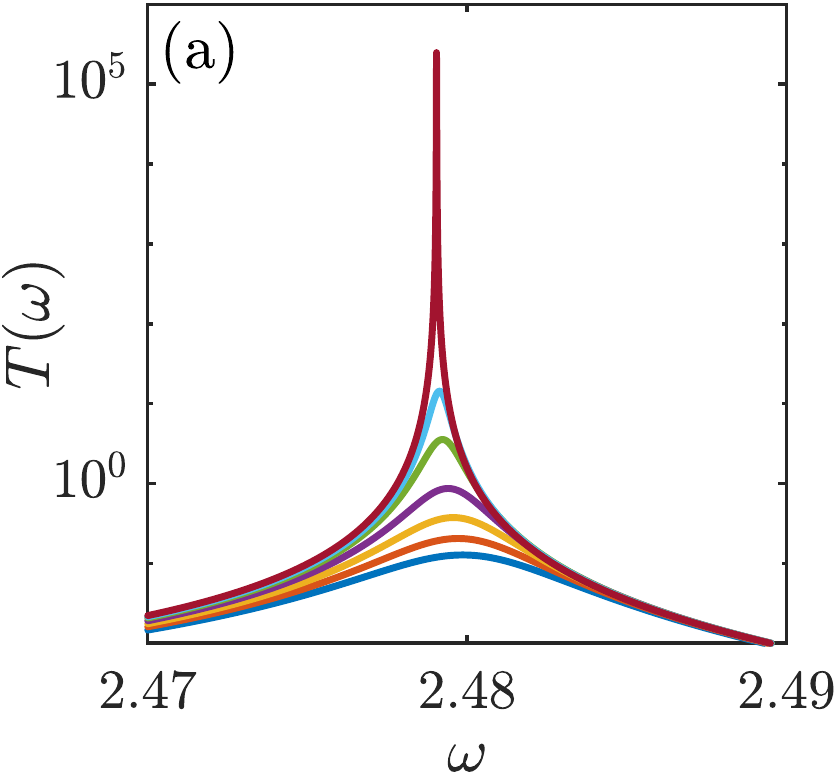}
		\hspace{-6pt}
		\includegraphics[valign=t,scale=0.5]{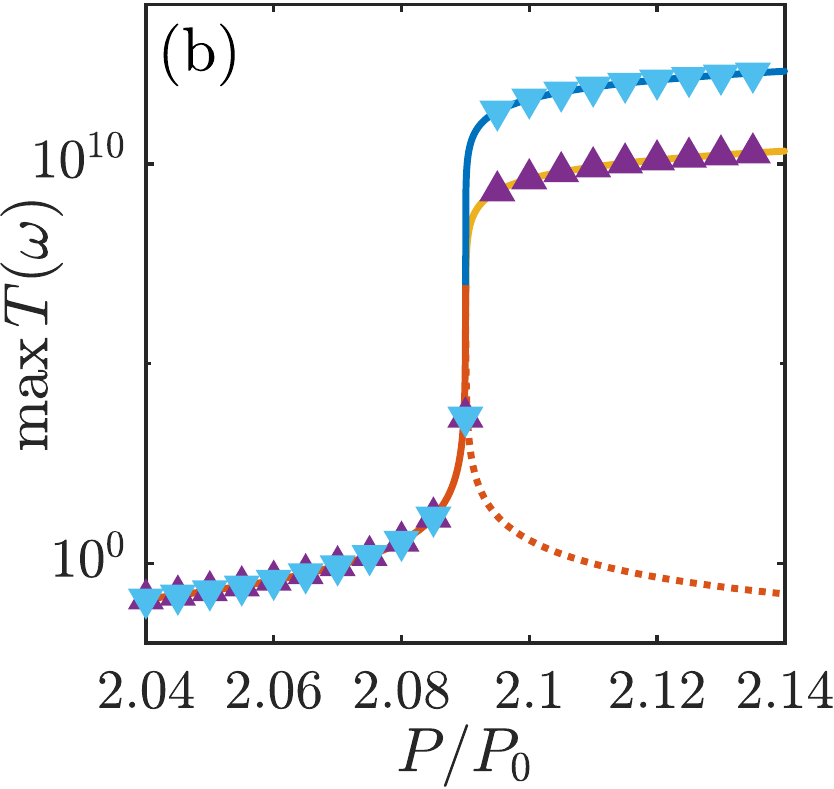}
	}
	\caption{Left panel (a): Incident-frequency-dependent transmission spectrum for a $1\times 7$ PEG and different gain strengths (from bottom to top) $P/P_0=2.04, 2.05, 2.06, 2.07, 2.08, 2.085, 2.09$ approaching the lasing threshold. Right panel (b): peak transmittivity as a function of gain strength for incident amplitude $E_0/\sqrt{J}=10^{-7}$ (upwards triangles) or $10^{-8}$ (downwards triangles). Red lines indicate the result of the linearized calculation based on the input-output formalism of ~\cite{Hafezi2011,Peano2016}.}
	\label{fig:08}
\end{figure}

A typical way to characterize the convective vs. absolute nature of a dynamical instability in generic nonlinear dynamical systems is to study the intensity distribution in the presence of some external noise and look for the so-called noise-sustained structures (NSS)~\cite{Cross1993,Santagiustina1997,Deissler1985,Louvergneaux2004}.

Generic quantum optical systems are unavoidably subject to quantum noise due to the discreteness of the light quanta.
An easy way to include the effects of the quantum noise is to switch to the Wigner representation \cite{Gardiner2000,Steel1998,Sinatra2002,Carusotto2005} and write stochastic differential equations for the classical complex variables $a_{m,n}$ corresponding to the quantum field amplitudes $\hat{a}_{m,n}$.
In the absence of extra noise sources, noise can be approximated by its expression in the linear gain regime, where it amounts to an additional stochastic term in \eqref{eq:EOM}:
\begin{equation}
	\dot{a}_{m,n}(t) =  \ldots
	-\sqrt{{\gamma}\left(1+\frac{P_{m,n}}{P}\right)}\,\xi_{m,n}(t).
	\label{eq:EOM_Wigner}
\end{equation}
Here the dots $\ldots$ summarize the RHS of \eqref{eq:EOM} and $\xi_{m,n}(t)$ are independent, zero-mean normally-distributed complex white noises of variance $1$.

Examples of simulations of noise-sustained structures in the topological laser PEG configuration are shown in Fig.~\ref{fig:06}. We go through the different instability regimes by varying the pump power $P$, namely $P/P_0 = 0.9$ for the absolutely stable (AS) regime, $P/P_0 = 1.5$ for the CI regime and $P/P_0 = 1.8$ for the AI regime. As shown in Fig.~\ref{fig:05}(b) for the considered $J/\gamma=10$ and $N=21$ case, the thresholds are at $P/P_0=1.13$ for the AS to CI transition and at $P/P_0=1.56$ for the CI to AI transition.

The spatio-temporal patterns in Fig.~\ref{fig:06} are calculated in three different cases, namely with a weak initial noisy seed (top row); with a coherent pulse incident on the system at a given time (central row); with a continuous white noise active during the whole evolution according to Eq.~\ref{eq:EOM_Wigner} (bottom row). 

In the CI regime without noise [panel (b)], the initial noisy seed gets quickly amplified in the amplifying region but is simultaneously advected away with group velocity $v_g$. Locally, the system then quickly returns to the equilibrium zero-field state. 
In the presence of continuous noise [panel (h)], the spatio-temporal pattern clearly shows so-called \emph{noise-sustained structures} (NSS)~\cite{Cross1993,Deissler1985}. These consist of high intensity stripes that keep appearing at random times and get amplified while being advected away at $\pm v_g$ (depending on which topological gap they are spectrally located in). As a result, the intensity is continuously fluctuating at all points, but its average and variance are strongest on the edge of the amplifying region. Of course the field shows no long-time phase coherence.
A short coherent pump pulse whose spectrum overlaps one of the two topological edge modes (the upwards propagating one in the figure) injects a wavepacket that gets quickly amplified while it propagates along the edge at $v_g$ with a minor spatial broadening [panel (e)]. Once the wavepacket reaches the edge of the amplified region, it starts decaying. 

These are the typical features of systems located in a convective instability regime and accurately match the ones displayed by other optical systems in the same regime~\cite{Santagiustina1997,Louvergneaux2004}.
For the sake of completeness, it is interesting to compare these behaviours to the ones in the absolute stability and in the absolute instability regimes.

In absence of noise [panels (a)-(c)] the initial perturbation gets quickly damped in the AS regime, while it is exponentially amplified into a self-supporting lasing mode in the AI regime. In this latter case, the chirality of the lasing mode is randomly selected depending on the initial condition. In the shown case, the system starts lasing in both chiralities, but eventually one of them (the upwards propagating one in the figure) dominates and ends up completely suppressing the other one.

We now replace the initial noisy perturbation with a short Gaussian pulse spectrally overlapping with the upwards propagating chiral edge mode. In the AS regime, we observe that the pulse propagates at $v_g$ but is quickly damped during propagation [panel (d)]. In the AI regime, instead, the injected pulse has the time to expand across the whole amplified region before being advected away, so that it can eventually transform into a self-supporting lasing mode [panel (f)]. In this case, the chirality of the lasing mode is fixed from the beginning by the one of the incident pulse.

In the presence of noise at all times, the stripe-shaped intensity fluctuations that are visible in the amplifying region have different properties in the AS regime [panel (g)] as compared to the one discussed above for the CI regime [panel (h)]. Since decay now dominates over amplification, the intensity is now roughly uniform across the whole amplified region and is no longer peaked on the edges. Still, both chiralities are randomly selected during the evolution.

In the AI regime, the behaviour in the presence of a continuous noise  [panel (i)] is very similar to the other two cases [panels (c) and (f)]. As in (c), the chirality of the lasing mode is randomly selected. The main difference with (c) and (f) is that the noise accelerates the onset of lasing; furthermore, weak intensity fluctuations are visible on top of the lasing mode at all times and propagate in the same direction.

\subsection{Robustness to disorder}

In order to assess the robustness of lasing to disorder, we now consider a PEG configuration with a $1 \times 15$ strip of amplifying sites on the left border of a $11 \times 51$ lattice and we add the typical Gaussian disorder configuration shown in Fig.~\ref{fig:07}(g). Snapshots of the spatial intensity distribution of the emission at a late time $t = 500\gamma^{-1}$ are shown in panels (d-f) for different values of the overall disorder strength. These plots suggest that the disorder strength which is needed to spoil the single mode nature of the topological laser emission is roughly 1/3 of what was needed in the WEG configuration discussed above. This relative fragility is due to proximity (visible in Fig.~\ref{fig:01}) of the lasing frequency to the bulk bands: a weaker disorder is sufficient to mix the edge mode with the bulk bands and thus break the edge state into independently lasing regions as shown in panel (f).

Further light on this physics can be obtained from the power spectral densities shown in panels (a-c). In contrast to the WEG case, no visible difference is found between the spectra for single realizations of lasing and the averaged ones. As already mentioned for the disorder-free case, this is due to the open boundaries of the amplifying region, which allow for a smooth adjustment of the lasing mode to the optimal gain.  As long as the disorder remains moderate, we have a monochromatic and single mode emission. For the strong disorder strength case considered in panel (c), the spatial breaking into several independent lasing mode visible in panel (f) reflects in the multi-mode character of the emission, which also involves frequencies located within the bands.

\subsection{Amplification of a propagating probe}

As a final characterization of the CI regime, Fig.~\ref{fig:08} illustrates the possibility of an efficient traveling-wave amplification~\cite{Peano2016}. We consider a system of $11 \times 25$ sites with amplification restricted to a $1\times 7$ vertical strip in the middle of the left border (sites 10 to 16). The chiral transmission of a coherent probe through the gain region is studied using a pair of input and output waveguides coupled to the neighboring sites 8 and 18 on the same border. 
The transmission is calculated by solving the temporal evolution until the steady state is reached. As usual in input-output theory~\cite{Gardiner1985}, new terms must be added to the time-evolution equations for the input and output sites, 
\begin{eqnarray}
	\dot{a}_{\mathrm{in}}(t) &=& \ldots - \frac{\gamma_{\mathrm{in}}}{2}a_{\mathrm{in}}- \sqrt{\gamma_{\mathrm{in}}}E_0e^{-i\omega t} \\
	\dot{a}_{\mathrm{out}}(t) &=& \ldots - \frac{\gamma_{\mathrm{out}}}{2}a_{\mathrm{out}}
	\label{eq:IO_num_modification}
\end{eqnarray}
where the dots $\ldots$ summarize the RHS of \eqref{eq:EOM}, the incident field has amplitude $E_0$ and frequency $\omega$, and $\gamma_{\mathrm{in,out}}$ account for the extra radiative losses into the waveguides. The transmittivity (Fig.~\ref{fig:03}) is obtained from the transmitted field $E_{\mathrm{out}} = \sqrt{\gamma_{\mathrm{out}}}a_{\mathrm{out}}$ as $T=|E_{\mathrm{out}}/E_0|^2$: below the lasing threshold $P_{\mathrm{abs}}$, the full numerical calculations (triangles) are perfectly recovered by a simpler linearized calculation based on the Green's function approach for a weak probe (red lines) discussed in the SM of~\citep{Hafezi2011} and extended to the quantum level in~\cite{Peano2016}. Above the threshold, nonlinear effects dominate and the linearized calculations are no longer reliable.

Panel (a) shows the transmission spectrum for gain values in the CI region $\tilde{P}_0<P<P_{\mathrm{abs}}$. For $P<P_0$, gain is not able to overcome losses: the net absorption of all sites combined with the impedance mismatch at the input and output waveguides conspire to give a very low transmission. As $P$ grows above $\tilde{P}_0$, net amplification sets in, giving a broad transmission peak. As $P$ further grows towards $P_{\mathrm{abs}}$, the transmittivity grows far above $1$ in a narrow frequency range and eventually diverges at the lasing frequency as the absolute threshold is approached ($P\to P_{\mathrm{abs}}^-$). Panel (b) shows the peak transmittivity as a function of gain strength for two values of the probe intensity. Well below the laser threshold, the two curves coincide as the system behaves in a linear way. Around and above threshold, instead, nonlinear gain saturation sets in, limiting the effective amplification and thus distinguishing the two curves. Well above the laser threshold, the field intensity is fixed by the self-oscillation process independently of the probe, so the transmittivity is inversely proportional to $|E_0|^2$.

\section{Conclusions}
\label{sec:conclusions}

In this article, we have reported a theoretical study of a topological laser device based on a bosonic Harper-Hofstadter lattice model with optical gain. Striking consequences of the chirality of the lasing mode have been highlighted: when gain is distributed around the whole edge, lasing can occur in a number of closely-spaced modes and relaxation towards the steady-state occurs on a very slow timescale; when gain is restricted to a finite strip, relaxation is fast but the distinction between convective and absolute instabilities causes an increase of the threshold and introduces new amplification regimes. To complete the picture, we have quantitatively assessed the impact of disorder on topological lasing and highlighted the stronger robustness of the WEG configuration. Finally, in analogy to other convectively unstable systems, we have illustrated the qualitative shape of the structures that appear in the presence of noise.

Future steps include the extension of our theory to specific models of the amplifying medium displaying a non-trivial carrier dynamics and a frequency-dependent gain, the development of a general theory of the collective excitations on top of a topological laser emission, and the construction of a quantum theory of topological lasing including quantum fluctuations. These will be crucial steps towards a complete understanding of the ultimate limits to the performance of topological laser devices.

\acknowledgments

Continuous exchanges with M. Wouters on the subject of convective and absolute instabilities are warmly acknowledged. We are grateful to T. Ozawa, H. M. Price and A. Loirette-Pelous for stimulating discussions.
I.C. acknowledges financial support from the Provincia Autonoma di Trento and from the European Union via the FET-Open Grant MIR-BOSE (737017) and Quantum Flagship Grant PhoQuS (820392).
M.C. acknowledges support from MIUR PRIN 2015 (Prot. 2015C5SEJJ001) and SISSA/CNR project ``Superconductivity, Ferroelectricity and Magnetism in bad metals'' (Prot. 232/2015).

\bibliographystyle{apsrev4-1}   
\bibliography{references}

\end{document}